

\def\today{\number\day\enspace
     \ifcase\month\or January\or February\or March\or April\or May\or
     June\or July\or August\or September\or October\or
     November\or December\fi \enspace\number\year}
\def\clock{\count0=\time \divide\count0 by 60
    \count1=\count0 \multiply\count1 by -60 \advance\count1 by \time
    \number\count0:\ifnum\count1<10{0\number\count1}\else\number\count1\fi}
\footline={\ifnum\pageno=1 \hss \hfill \hss
           \else \hss\rm -- \folio\ -- \hss\fi}

\overfullrule=0pt 
\raggedbottom
\def\push#1{\hbox to 1truein{\hfill #1}}


\def\arcsec{\ifmmode^{\prime\prime}\else $^{\prime\prime}$\fi}
\def\arcmin{\ifmmode^{\prime}\else $^{\prime}$\fi}
\def\deg{\ifmmode^\circ\else$^\circ$\fi}
\def\sun{\ifmmode_{\mathord\odot}\else$_{\mathord\odot}$\fi}

\def\Teff{{\it T}\lower.5ex\hbox{\rm eff}}
\def\kms{\ifmmode{{\rm km~s}^{-1}}\else{km~s$^{-1}$}\fi}
\def\pervol{\ifmmode{{\rm cm}^{-3}}\else{cm$^{-3}$}\fi}
\def\perarea{\ifmmode{{\rm cm}^{-2}}\else{cm$^{-2}$}\fi}
\def\mum{\ifmmode{\mu {\rm m}}\else{$\mu$m}\fi}
\def\respwr{\ifmmode{\lambda /\Delta\lambda}
    \else{$\lambda /\Delta\lambda$}\fi}
\def\whz{\ifmmode{{\rm W~Hz}^{-1/2}}\else{W~Hz$^{-1/2}$}\fi}
\def\wig#1{\mathrel{\hbox{\hbox to 0pt{%
    \lower.5ex\hbox{$\sim$}\hss}\raise.4ex\hbox{$#1$}}}}
\def\lsim{\wig <}
\def\gsim{\wig >}
\def\sqr#1#2{{\vcenter{\vbox{\hrule height.#2pt
    \hbox{\vrule width.#2pt height#1pt \kern#1pt
    \vrule width.#2pt}
    \hrule height.#2pt}}}}

\def\boxit#1{\vbox{\hrule\hbox{\vrule\kern3pt
   \vbox{\kern3pt#1\kern3pt}\kern3pt\vrule}\hrule}}


\def\etal{{\it et~al.\/}}

\def\eg{{\it e.g.\/}}
\def\cf{{\it cf.\/}}
\def\ie{{\it i.e.\/}}

\hyphenation{nucleo-cos-mo-chron-ology}  
\hyphenation{Nucleo-cos-mo-chron-ology}
\hyphenation{chron-o-meter}
\hyphenation{Ro-ches-ter}
\hyphenation{mol-e-cules}



\def\Syn6301{{\em Synechococcus\/} 6301}


\def\jref#1 #2 #3 #4 {{\par\noindent \hangindent=3em \hangafter=1
      \advance \rightskip by 5em #1, {\it#2}, {\bf#3}, #4.\par}}
\def\ref#1{{\par\noindent \hangindent=3em \hangafter=1
      \advance \rightskip by 5em #1\par}}


\newcount\eqnum
\def\nexteq{\global\advance\eqnum by1 \eqno(\number\eqnum)}
\def\lasteq#1{\if)#1[\number\eqnum]\else(\number\eqnum)\fi#1}
\def\preveq#1#2{{\advance\eqnum by-#1
    \if)#2[\number\eqnum]\else(\number\eqnum)\fi}#2}


\def\endtable{\endgroup}
\def\tableheight{\vrule width 0pt height 8.5pt depth 3.5pt}
{\catcode`|=\active \catcode`&=\active
    \gdef\tabledelim{\catcode`|=\active \let|=\vbar
                     \catcode`&=\active \let&=\nobar} }
\def\table{\begingroup
    \def\twidth{\hsize}
    \def\tablewidth##1{\def\twidth{##1}}
    \def\defaultheight{\vrule width 0pt height 8.5pt depth 3.5pt}
    \def\heightdepth##1{\dimen0=##1
        \ifdim\dimen0>5pt
            \divide\dimen0 by 2 \advance\dimen0 by 2.5pt
            \dimen1=\dimen0 \advance\dimen1 by -5pt
            \vrule width 0pt height \the\dimen0  depth \the\dimen1
        \else  \divide\dimen0 by 2
            \vrule width 0pt height \the\dimen0  depth \the\dimen0 \fi}
    \def\spacing##1{\def\defaultheight{\heightdepth{##1}}}
    \def\nextheight##1{\noalign{\gdef\tableheight{\heightdepth{##1}}}}
    \def\end{\cr\noalign{\gdef\tableheight{\defaultheight}}}
    \def\zerowidth##1{\omit\hidewidth ##1 \hidewidth}
    \def\hline{\noalign{\hrule}}
    \def\skip##1{\noalign{\vskip##1}}
    \def\bskip##1{\noalign{\hbox to \twidth{\vrule height##1 depth 0pt \hfil
        \vrule height##1 depth 0pt}}}
    \def\header##1{\noalign{\hbox to \twidth{\hfil ##1 \unskip\hfil}}}
    \def\bheader##1{\noalign{\hbox to \twidth{\vrule\hfil ##1
        \unskip\hfil\vrule}}}
    \def\spanloop{\span\omit \advance\mscount by -1}
    \def\extend##1##2{\omit
        \mscount=##1 \multiply\mscount by 2 \advance\mscount by -1
        \loop\ifnum\mscount>1 \spanloop\repeat \ \hfil ##2 \unskip\hfil}
    \def\vbar{&\vrule&}
    \def\nobar{&&}
    \def\hdash##1{ \noalign{ \relax \gdef\tableheight{\heightdepth{0pt}}
        \toks0={} \count0=1 \count1=0 \putout##1\end
        \toks0=\expandafter{\the\toks0 &\end} \xdef\piggy{\the\toks0} }
        \piggy}
    \let\e=\expandafter
    \def\putspace{\ifnum\count0>1 \advance\count0 by -1
        \toks0=\e\e\e{\the\e\toks0\e&\e\multispan\e{\the\count0}\hfill}
        \fi \count0=0 }
    \def\putrule{\ifnum\count1>0 \advance\count1 by 1
 
\toks0=\e\e\e{\the\e\toks0\e&\e\multispan\e{\the\count1}\leaders\hrule
        \hfill}
        \fi \count1=0 }
    \def\putout##1{\ifx##1\end \putspace \putrule \let\next=\relax
        \else \let\next=\putout
            \ifx##1- \advance\count1 by 2 \putspace
            \else    \advance\count0 by 2 \putrule \fi \fi \next}   }
\def\tablespec#1{
    \def\vdimens{\noexpand\tableheight}
    \def\tabby{\tabskip=0pt plus100pt minus100pt}
    \def\r{&################\tabby&\hfil################\unskip}
    \def\c{&################\tabby&\hfil################\unskip\hfil}
    \def\l{&################\tabby&################\unskip\hfil}
    \edef\templ{\noexpand\vdimens ########\unskip  #1
         \unskip&########\tabskip=0pt&########\cr}
    \tabledelim
    \edef\body##1{ \vbox{
        \tabskip=0pt \offinterlineskip
        \halign to \twidth {\templ ##1}}}
    \edef\sbody##1{ {
        \tabskip=0pt \offinterlineskip
        \halign to \twidth {\templ ##1}}}
}


\def\ltextindent#1{\indent\llap{\hbox to \parindent {#1\hfil}}\ignorespaces}

\def\mathfont#1{{
    #1\count20=\fam\multiply\count20 by "100\advance\count20 by "7000
    \count21=`a \advance\count21 by - 1
    \count22=\count21\advance\count22 by \count20
    \loop \advance\count22 by 1 \advance\count21 by 1
             \global\mathcode\count21=\count22
    \ifnum \count21<`z \repeat
    \count21=`A \advance\count21 by - 1
    \count22=\count21\advance\count22 by \count20
    \loop \advance\count22 by 1 \advance\count21 by 1
             \global\mathcode\count21=\count22
    \ifnum \count21<`Z \repeat}}

\outer\def\section#1\par{\vskip 12pt plus 10pt minus 4pt
    \centerline{\bf#1}\nobreak\vskip 5pt plus 3pt minus 2pt}


\font\title=cmbx10 scaled \magstep2

\def\Teff{{T_{\rm eff}}} 

\def\eg{{\it e.g.}}
\def\cf{{\it cf.}}
\def\ie{{\it i.e.}}
\def\arcsec{\hbox{$^{\prime\prime}$}}
\def\Lsun{L_{\odot}}  
\def\Msun{M_{\odot}}  
\def\Rsun{R_{\odot}}  


\noindent agbppr3.tex, version of 11 March 2002 - revised by JHT

\vskip 10pt

\noindent{\bf FUEL-SUPPLY-LIMITED STELLAR RELAXATION OSCILLATIONS:
APPLICATION TO MULTIPLE RINGS AROUND AGB STARS AND PLANETARY
NEBULAE}

\vskip 10pt

\centerline{Hugh M. Van Horn\footnote{$^1$}{Current address: Department of
Terrestrial Magnetism, Carnegie Institution of Washington, 5241 Broad
Branch
Road, Washington, DC 20015; on leave from the National Science
Foundation},
John H. Thomas\footnote{$^2$} {Also Department of Mechanical Engineering,
University of Rochester}, Adam Frank, and Eric G. Blackman}

\vskip 10pt

\centerline{Department of Physics and Astronomy}

\centerline{University of Rochester}

\centerline{Rochester, NY 14627-0171}

\vskip 10pt

\centerline{\bf Abstract}

\vskip 10pt

    We describe a new mechanism for pulsations in evolved stars: relaxation 
oscillations driven by a coupling between the luminosity-dependent mass-loss
rate and the H fuel abundance in a nuclear-burning shell.  
When mass loss is
 
included, the outward flow of matter can modulate the flow of fuel into the 
shell when the stellar luminosity is close to the Eddington luminosity
$L_{\rm Edd}$.  When the luminosity drops below $L_{\rm Edd}$, the mass outflow 
declines and the shell is re-supplied with fuel.  This process can be 
repetitive.  We demonstrate the existence of such oscillations and discuss
the dependence of the results on the stellar parameters.  In particular, we show
that the oscillation period scales specifically with the mass of 
the H-burning relaxation shell (HBRS), 
defined as the part of the H-burning shell above 
the minimum radius at which the luminosity from below first exceeds the 
Eddington threshold at the onset of the mass loss phase.
For a stellar mass $M_*\sim 0.7\Msun$,
luminosity $L_*\sim 10^4\Lsun$, and mass loss rate $|\dot M|\sim
10^{-5}\Msun$ yr$^{-1}$, the oscillations have a recurrence time $\sim 1400$
years $\sim 57\tau_{\rm fsm}$, where $\tau_{\rm fsm}$ is the timescale for 
modulation of the fuel supply in the HBRS 
by the varying mass-loss 
rate.  This period agrees very well with the $\sim$ 1400-year period inferred for 
the spacings between the shells surrounding some planetary nebulae.  We also
find the half-width of the luminosity peak to be $\sim 0.39$ times the 
oscillation period; for comparison, the observational shell thickness of 
$\sim$ 1000 AU corresponds to $\sim 0.36$ of the spacing between pulses.
We find  oscillations only for models in which the luminosity of the 
relaxation shell is $\sim$ 10--15\% of the total stellar luminosity 
and for which energy 
generation occurs through the $pp$ chain.  We suggest this mechanism as a 
natural explanation for the circumnebular shells surrounding some planetary 
nebulae, which appear only at the end of the AGB phase. 

\vskip 10pt

\noindent{\bf 1. Introduction}

\vskip 10pt

    Asymptotic giant branch (AGB) stars exhibit pulsations on a variety of 
timescales, ranging from long periods associated with thermal pulses of the 
He-burning shells ($\tau \sim 10^5$ years; \cf,  Iben and Renzini 1983, 
Sch\"onberner 1983, Mazzitelli and D'Antona 1986, Iben 1991, Dorman \etal, 
1993, Vassiliadis and Wood 1993) to far shorter oscillations associated with
acoustic modes of the star ($\tau \sim 1$ year).  Recently, a new class of 
pulsational behavior has been inferred for such stars.  The evidence takes
the form of nested circumnebular shells around a number of planetary nebulae
(\cf, Mauron and Huggins 2000; Terzian and Hajian 2000; Balick, Wilson, and Hajian
2001).  These appear to have been ejected in pulses during the AGB phase or 
succeeding post-AGB but pre-planetary-nebula (pre-PN) phases of stellar 
evolution.  For example, surrounding the Cat's Eye Nebula (NGC 6543), Balick
\etal\ (2001) find rings that appear to have been ejected
quasi-periodically, with the estimated time between successive pulses 
ranging from $\sim$ 1000 to $\sim$ 1900 years; the mean spacing for eight 
successive shells is approximately 1400 years.  The authors take the rings 
to be $\sim$ 1000 AU thick, indicating that the ejection mechanism operates 
during 0.36 of the time between successive pulses.  Each shell is estimated 
to contain $\sim 0.01\Msun$ of material, and a total mass $\sim 0.1\Msun$ 
appears to have been ejected in the process that produced the shells. 

    To date, four models have been proposed to explain these enigmatic
shells.  
Soker and Harpez (1988) suggested a binary model, in which the secondary 
follows a highly elliptical orbit.  As the secondary approaches periastron,
it interrupts the mass loss from the primary, leading to the formation of a 
shell.  The intershell timescale is then set by the orbital period 
$\tau_{\rm orb}$ of the binary.  Mastrodemos and Morris (1998, 1999)
proposed a different binary model, in which the shells are the result of spiral
shocks in the AGB wind.  In these models the intershell timescale is also
$\tau_{\rm 
orb}$.  Simis, Icke, and Dominik (2001) proposed that the shells are
produced by quasi-periodic variations in dust formation rates in dust-driven AGB
winds.  They argue that non-linear coupling between pulsation of the star, dust 
formation in the atmosphere, and radiation-driving in the wind produces
shells with a quasi-regular period $\sim 10^3$ years.  Finally, Garcia-Segura
\etal\ (2002) have proposed an MHD model in which a dynamo operating in the star
sets the period of the oscillations, and magnetic pressure in the wind creates
the shells.  It is difficult at present to evaluate the validity of these
models.  
The eccentric binary model seems unlikely to account for the several PN 
discovered with multiple shells; it is not clear why a $\sim 10^3$ year
period should be favored; and it is hard to understand how the observed variation 
from $\sim$ 1000 to 2000 years in shell spacings could be produced by a 
perfectly periodic binary model.  The MHD model may be based on sound
physics, but existing AGB dynamo models (Blackman \etal\ 2001) predict
far shorter dynamo periods. 

    A number of indications seem to us to point instead to
fuel-supply-limited 
relaxation oscillations as the explanation for the observed circumnebular
mass shells: 

    (1) At high luminosities, stars lose mass at exceptionally high rates. 
Indeed, the brightest precursors of the planetary nebulae may have $L_*\gsim
L_{\rm Edd}$, driving the so-called \lq\lq superwind\rq\rq\ that terminates 
evolution up the AGB.  However, the effect of the stellar wind upon the 
internal structure of a star has not, to our knowledge, been included 
self-consistently in previous calculations.  Models of the wind assume that 
the stellar interior is dynamically unaffected by the mass flux through the 
surface layers.  Conversely, models of stellar evolution that include wind 
mass loss simply strip mass away from the stellar surface.  Because the rate
of mass loss increases dramatically as a star climbs the AGB (\cf,
Vassiliadis 
and Wood 1993), any effect due to the wind will be accentuated near the end
of 
the AGB phase. As the observed circumnebular shells were evidently produced 
during the final stages of evolution just prior to the ejection of a
planetary 
nebula, some effect of mass loss upon the internal stellar structure thus 
seems a natural candidate to explain the pulsed ejection of the
circumnebular 
shells. 

    (2) Comparison of the rate at which H burning consumes mass with the
rate 
of mass loss from the stellar surface also is suggestive. Let $Q^{\prime}$
be 
the amount of energy released per gram of H consumed.  This quantity is
(\cf, 
Clayton 1968, p. 287 ff) 

$$
Q^{\prime}={(4m_{\rm H}-m_{\rm He})c^2\over 4m_{\rm H}} =
6.398\times
10^{18}\
{\rm erg\ g^{-1}}.\eqno(1)
$$

\noindent If $X$ is the mass fraction of H in the unburned material, $\dot
M_s$ the rate at which an H-burning shell consumes mass, and $L_s$ the
resulting luminosity produced by this shell, then

$$
\dot M_s X =-{L_s\over Q^{\prime}}.\eqno(2)
$$

\noindent With $L_s\sim 10^4\Lsun$ and $X\sim 0.7$, the rate at which mass
must be consumed by a H-burning shell in order to provide the stellar luminosity
is thus $|\dot M_s|\sim 10^{-7}\Msun$ yr$^{-1}$.  For comparison, the observed 
rates of mass loss range from $5\times 10^{-11}\lsim (|\dot M|/\Msun\ {\rm 
yr}^{-1})\lsim 2\times 10^{-6}$ for the central stars of the planetary
nebulae 
(Perinotto 1989) to $\sim 10^{-4}\Msun$ yr$^{-1}$ for long-period variables 
(\cf, Cox \etal\ 2000, p. 415).  We note that the OH/IR stars which have the
highest mass loss rates may have $L_*\gsim L_{\rm Edd}$.  Thus, at most
stages 
in an AGB star's evolution, mass loss from the surface is comparable to or 
higher than mass consumption at the H-burning shell.  Consequently, there is
a competition between the need to supply fuel downward into the H-burning
shell 
and upward into the stellar wind. 
    
    We are thus led to suggest that an interaction between radiation-driven 
mass loss from the stellar surface and mass consumption by 
H-burning 
in a highly evolved AGB star may produce a relaxation oscillation that
ejects 
shells of matter quasi-periodically.  Near the tip of the AGB, a star has a 
luminosity $L_*\sim 10^4\Lsun$, and it contains a dense, degenerate C/O core
with mass $\sim 0.6\Msun$ and radius $\sim 0.02 \Rsun\sim 10^9$ cm,
surrounded 
by a tenuous H/He envelope extending outward to radius $R_*\sim 4\times 
10^{13}\ {\rm cm}\sim 500\Rsun$ (\cf, Iben 1987).  Suppose that such a star
is 
initially in a steady state.
We define the H-burning relaxation shell (HBRS) 
as a thin outer part of the 
H-burning region above the smallest radius for which 
the luminosity from below can exceed the Eddington luminosity.
This region can be quite a small fraction of the total H-burning
shell, but is the region which is important for relaxation oscillations
that we now discuss. 

We assume that the HBRS consumes mass at a 
rate $|\dot M_s|$.  Suppose also that the star undergoes radiation-driven
mass 
loss from the stellar surface at a steady rate $| \dot M |$.  A perturbation
that increases the temperature of the HBRS, leading to a 
thermonuclear runaway, greatly increases the luminosity.  Because the
surface 
mass-loss rate is proportional to the luminosity (\cf, \S 2), the increased 
luminosity increases the rate of mass loss from the surface.  Conservation
of mass, however, requires that the mass flowing out through the surface must
be replaced from below; that is, mass loss is not simply a stripping away of the
outer layers of the star.  Some form of gradual readjustment must occur
along the AGB and into the post-AGB and pre-PN phases.  The increased mass-loss
rate thus at least impedes the supply of fuel to the HBRS, and it may 
even temporarily interrupt that supply.  In any event, the increased surface
mass loss has the effect of reducing the luminosity produced by the HBRS, 
which in turn reduces the rate of radiation-driven mass loss from the
stellar surface, allowing the cycle to repeat itself.  This is a previously 
unexplored coupling between the HBRS luminosity, the mass-loss
rate, and the fuel supply available in the HBRS.  Such 
fuel-supply-limited relaxation oscillations are familiar in other contexts, 
such as the pulsation of a fuel-starved flame from a guttering candle. 

    In the present paper, we consider in more detail the hypothesis that the
source of the circumnebular shells is just such a fuel-supply-limited 
relaxation oscillation in a star near the end of its AGB phase.  In \S 2, we
first review briefly the relation between the rate of mass loss $\dot M$ and
the stellar luminosity $L_*$ for the optically thick, radiation-driven winds
appropriate to high-luminosity stars.  In \S 3, we next collect together the
fundamental Eulerian equations that describe the hydrodynamic interaction 
between matter and radiation in a spherically symmetric star with a 
spherically symmetric mass outflow.  In \S 4 we develop a simplified model 
that accounts qualitatively for the main features of the relaxation 
oscillations described above.  We emphasize that this is a simplified model,
intended to illustrate the essential physics of these oscillations. A more 
complete calculation, beyond the scope of this paper, would require a
detailed 
stellar evolution model including self-consistent mass loss. We describe the
results of numerical calculations using our simplified model in \S 5, and we
conclude in \S 6 with a summary. 

\vskip 10pt

\noindent{\bf 2. Mass Loss on the AGB}

\vskip 10pt

    From his observations of mass loss from luminous K and M giants and 
supergiants, Reimers (1975a, b) proposed an empirical formula to express the
dependence of the rate of mass loss upon the stellar parameters: 

$$
|\dot M|= 4\times 10^{-13}\Msun\ {\rm
yr}^{-1}\cdot\eta{(L/\Lsun)(R/\Rsun)\over
(M/\Msun)},\eqno(3)
$$

\noindent where $\eta$ is a parameter of order unity.  This expression has 
been used extensively, both in analyzing observations and in theoretical 
studies. For example, for an AGB star with $M_*\sim \Msun$, $L_*\sim 
10^4\Lsun$, and $R_*\sim 500\Rsun$, this expression yields the result $|\dot
M|\sim 2\times 10^{-6}\Msun$ yr$^{-1}$.  However, as originally pointed out
by 
Renzini (1981; see also Iben and Renzini 1983), this equation appears to 
underestimate the rate of mass loss in the very high luminosity, thermally 
pulsing AGB phase, requiring a postulated \lq\lq superwind\rq\rq\ to provide
the necessary mass-loss rates. 

    Bowen (1988) and Bowen and Willson (1991) have described hydrodynamic 
calculations of the atmospheres of long-period, Mira variables -- pulsating 
AGB stars of low to intermediate masses -- that lead to significant rates of
mass loss.  In models that include dust formation, radiation pressure acting
on the dust grains produces a mass loss extending up to $\dot M\sim 10^{-
6}\Msun$ yr$^{-1}$ for a star with $M_*=1.2\Msun$, $R_*=270\Rsun$, and 
$L_*=5315\Lsun$. Bowen suggests this as the \lq\lq superwind\rq\rq\ that 
appears to terminate AGB evolution.  Willson (2000) has recently 
re-interpreted the empirical Reimers mass-loss rate (3) in terms of these 
results. 

    In their computations of the evolution of intermediate-mass stars, 
Vassiliadis and Wood (1993) included an empirical mass-loss rate that 
reproduces the very large values of $\dot M$ observed during the AGB phase
of evolution.  It reaches an essentially constant \lq\lq superwind\rq\rq\ 
rate that is within a factor of two of the value 

$$
|\dot M|={L_*\over cv_{\rm w}}\sim 10^{-5}\Msun\ {\rm yr}^{-1}\times
{L_*\over
10^4\Lsun}\eqno(4)
$$

\noindent appropriate for a radiation-driven wind, where $v_{\rm w}\sim 15$
km 
s$^{-1}$ is the asymptotic wind speed far from the central star.  For stars 
with $M\lsim 2.5\Msun$, the \lq\lq superwind\rq\rq\ mass-loss rate is
achieved 
only during the last few thermal pulses, and the vast majority of the mass 
loss occurs in these episodes.  

    Koesterke, Dreizler, and Rauch (1998) and Koesterke and Werner (1998) 
measured the mass-loss rates for four PG 1159 stars -- post-AGB stars that 
are among the immediate precursors of the white dwarfs -- finding $-8\lsim 
\log(|\dot M|/\Msun\ {\rm yr}^{-1})\lsim -7$, consistent with the theory of 
radiation-driven winds. 

\vskip 10pt

\noindent{\bf 3. Fundamental Equations}

\vskip 10pt

    The fundamental Eulerian equations that describe the structure and time 
variation of a spherically symmetric star with spherically symmetric mass
loss 
are (\cf, Landau and Lifshitz 1959; Hansen and Kawaler 1994) the continuity 
equation, 

$$
{\partial\rho\over\partial t}+{1\over r^2}{\partial(r^2\rho
v)\over
\partial
r}=0,\eqno(5a)
$$

\noindent where $\rho$ is the density and $v$ is the radial velocity; the
definition of the mass $M_r$ interior to radius $r$,

$$
{\partial M_r\over \partial r}=4\pi r^2\rho;\eqno(5b)
$$

\noindent the equation of conservation of radial momentum,

$$
{\partial P\over \partial r} + \rho\left({\partial v\over\partial
t}
+v{\partial v\over\partial r}\right) = -\rho{GM_r\over
r^2},\eqno(5c)
$$

\noindent where $P=P_g+P_r$ is the sum of the gas pressure $P_g=\rho k_B
T/\mu 
H$ and radiation pressure $P_r={1\over 3}a_r T^4$, and where $T$ is the 
temperature; and the heat-flow equation 

\def\sd{\strut\displaystyle}  

$$
{\partial T\over \partial r}=\cases{-{\sd 3\over \sd 4ac}{\sd
\kappa\rho\over
\sd T^3}{\sd L_r\over \sd 4\pi r^2},&in radiative equilibrium;\cr
    \nabla_{\rm ad}{\sd T\over \sd P}{\sd \partial P\over \sd
\partial
r},&in
convective equilibrium.}\eqno(5d)
$$

\noindent Here $\kappa$ is the opacity of the stellar matter, and $L_r$ is
the 
luminosity flowing out through a sphere of radius $r$.  Note that the first 
form of equation (5d) assumes that energy transport is dominated by
radiation 
flow rather than convection, while the second form is that appropriate for 
convective energy transport.  We also require the equation of energy 
conservation, 

$$
{\partial L_r\over\partial r}=4\pi r^2\rho\left(\epsilon -
T{\partial
S\over
\partial t} - Tv{\partial S\over\partial r}\right),\eqno(5e)
$$

\noindent where $\epsilon$ is the thermonuclear energy generation rate, and 
$S$ is the entropy.  In addition, we need the equation for the time rate of 
change of the H mass fraction $X$ due to nuclear burning in the HBRS: 

$$
{\partial X\over \partial t}+v{\partial X\over \partial r}
=-{\epsilon\over
Q^{\prime}},\eqno(5f)
$$

\noindent where $Q^{\prime}$ is given by equation (1).  Note that if we 
multiply equation (5f) by $dM_r$ and integrate over the HBRS, we 
recover equation (2), with $\dot M_s\equiv 4\pi r_s^2\rho_sv_s$. 

\vskip 10pt

\noindent{\bf 4. A Simplified Model}

\vskip 10pt

    Integrating equation (5b) from the center of the star out to some {\it 
fixed} radius $r$ and computing the time rate of change of this quantity
gives 

$$
\dot M_r\equiv{dM_r\over dt}=\int_0^r 4\pi r^2{\partial\rho\over \partial
t}dr
= -4\pi r^2\rho v\equiv -F(r, t),\eqno(6)
$$

\noindent where we use equation (5a) to eliminate $\partial\rho/\partial t$,
and the last equivalence {\it defines} the local mass-flow rate $F$.  At the
stellar surface, $\dot M_r\rightarrow\dot M=-F(R_*[t],t)$, a quantity that 
depends only upon time.  If the star experiences steady-state mass loss,
then 
equation (5a) shows that the mass flux $F(r,t)$ is a constant, independent
of 
position as well as time.  In general, $F$ may depend upon both position and
time, but if the changes in the star occur quasi-statically (as is the case
in stellar evolution calculations), then we can expect the mass flux through
the star to remain close to its (position-independent) value throughout the
entire stellar envelope. 

    From a scale analysis of the momentum equation (5c), we find that the
ratio 
of the velocity-dependent terms to the remaining terms is of order 
$(v/c_s)^2$, where $c_s$ is the sound speed.  For a mass-loss rate of $10^{-
5}\Msun$ yr$^{-1}$, and assuming the mass flow rate $F\equiv 4\pi r^2\rho v$
to be independent of position in the stellar envelope, the flow velocity at 
the HBRS is $v\sim 0.5$ cm s$^{-1}$.  In comparison, the sound 
speed at the shell is $c_s\sim 4\times 10^7$ cm $^{-1}$.  Thus, at the 
HBRS, the velocity-dependent terms in equation (5c) are only
$\sim 
10^{-16}$ times the other terms in this equation.  These terms also appear
to 
be negligible near the photosphere, and for this reason, we neglect them in 
the momentum equation.  This approximation is equivalent to assuming that
the 
variations of interest to us here all take place on timescales long in 
comparison with the hydrostatic readjustment timescale.  For this reason, 
the hydrodynamic timescale $\tau_{\rm hyd}\sim R_*/v \sim 500\Rsun/15$ km 
s$^{-1} \sim 1$ year plays no further role in our development of the
equations 
that describe this model. 

    Dropping these $v$-dependent terms reduces the momentum equation to the 
usual equation of hydrostatic equilibrium.  Separating out the radiation 
pressure $P_r$, we find that the equation of hydrostatic equilibrium for the
gas pressure $P_g$ can be written as 

$$
{\partial P_g\over \partial r}=-{\rho GM_r\over r^2}\left[1-{\kappa
L_r\over
4\pi c GM_r}\right].\eqno(7)
$$

\noindent In the envelope of the star we can reasonably assume $L_r\approx
L_*={\rm constant}$ and $M_r\approx M_*={\rm constant}$.  With these
approximations, equations (7) and (5d) can be written in the forms

$$
{\partial P_g\over \partial r}=-{\rho GM_*\over r^2}\left[1-{\kappa
L_*\over
4\pi c GM_*}\right]\equiv -{\rho GM_*\over r^2}(1-\lambda),\eqno(8a)
$$

\noindent where the final equivalence defines the parameter $\lambda$, and
we 
have assumed radiative energy transport, at least near the HBRS: 

$$
{\partial T\over \partial r}=-{3\over 4a_rc}{\kappa\rho\over T^3}{L_*\over
4\pi r^2}.\eqno(8b)
$$

\noindent The quantity $\lambda$ in equation (8a) has a simple physical
interpretation -- it is a dimensionless form of the stellar luminosity:

$$
\lambda\equiv{\kappa L_*\over 4\pi cGM_*}\equiv {L_*\over L_{\rm
Edd}}.\eqno(9)
$$

\noindent where $L_{\rm Edd}\equiv 4\pi cGM_*/\kappa$ is the Eddington
luminosity.

    The assumption that radiative transport dominates in the stellar
envelope 
is not clearly justified {\it a priori}, as the highly distended surface 
layers of AGB stars are well known to contain deep convection zones.  We 
nevertheless believe that equation (8b) is appropriate here for at least two
reasons:  (1) The surface convection zone in an AGB star does not extend all
the way down to the HBRS.  We are most concerned with conditions 
near this shell, and it is well known that the temperature profile below the
convection zone in the surface layers of a star rapidly approaches the
\lq\lq 
radiative zero\rq\rq\ approximation we use below.  (2) It is not clear
whether 
the relaxation oscillations we wish to study actually occur on the AGB or 
whether they occur at some later high-luminosity phase, where the surface 
convection zone may be shallower and equation (8b) correspondingly better 
justified. 

    Dividing equation (8a) by equation (8b), assuming a Kramers' law opacity
of the form\footnote{$^*$}{The more familiar form of this equation is 
$\kappa=\kappa_0\rho T^{-3.5}$.  The form given in equation (10) is more 
convenient for our present purposes, however, and the two forms are
equivalent 
if $a=1$ and $b=1.125$.} 
    
$$ 
\kappa=\kappa_*P_g^a P_r^{-b},\eqno(10) 
$$ 
    
\noindent where $\kappa_*$, $a$, and $b$ are constants, and employing the 
conventional definition 
    
$$ 
P_g/P_r=\beta/(1-\beta)\eqno(11) 
$$ 
    
\noindent enables us to write the equation of hydrostatic equilibrium in a 
form that can be integrated directly, if we assume $\beta$ to be independent
of position in the star.  Using the \lq\lq radiative zero\rq\rq\ surface 
boundary condition, $P_g=0$ at $r=R_*$, we obtain 
    
$$ 
\lambda={(1-\beta)\over(b-a+1)}.\eqno(12) 
$$ 

\noindent Note that equation (12) includes the full effect of the Kramers'
law 
opacity, subject only to the approximation that $\beta$ is independent of 
position. 

    Using equation (10) for $\kappa$ in equation (8b), expressing $\rho$ in 
terms of $P_g$ and $T$ with the ideal gas law, and again using equation (11)
with $\beta$ assumed independent of position, we can also integrate equation
(8b), obtaining for the temperature distribution in the stellar envelope 
    
$$ 
T={(b-a+{1\over 4})\over (b-a+1)}{GM_*\mu H\over k}\left({1\over r}-{1\over 
R_*}\right),\eqno(13) 
$$ 
    
\noindent where we assume $T=0$ at $r=R_*$, consistent with the
approximation 
made in deriving equation (12).  Equation (13) expresses the temperature at 
position $r$ in terms of constants and stellar properties, including the 
stellar radius $R_*$. 

    Equations (12) and (13) express the conditions of hydrostatic and
thermal 
equilibrium in the stellar envelope.  Using these equations, we can also 
obtain an expression for the mass contained in the stellar envelope above
the 
HBRS at radius $r_s$: 

$$
\Delta M_{\rm env}\equiv\int_{r_s}^{R_*}4\pi \rho r^2dr =
    {\mu H\over k}{\beta\over(1-\beta)}{4\pi\over 3}R_*^3 a_r
    \left[\left({b-a+{1\over 4}\over b-a+1}\right){GM_*\mu H\over
kR_*}\right]^3
    \int_{r_s/R_*}^1\left({1\over s}-1\right)^3s^2ds,\eqno(14)
$$

\noindent where $s\equiv r/R_*$.  The integral in equation (14) can be
evaluated analytically and yields 

$$
\int_{r_s/R_*}^1\left({1\over s}-1\right)^3s^2ds = -\ln{r_s\over R_*} -
3\left(1-{r_s\over R_*}\right) +{3\over 2}\left[1-\left({r_s\over
R_*}\right)^2\right] - {1\over 3}\left[1-\left({r_s\over
R_*}\right)^3\right].\eqno(15)
$$

\noindent Note that the factors of $R_*$ cancel in the constants in equation
(14), so that the envelope mass depends only upon $\beta$ (and thus, from 
equation [12] upon $\lambda$), $M_*$, and the fractional radius $r_s/R_*$. 
Consequently, as the stellar luminosity increases ($\lambda$ increases and 
$\beta$ accordingly decreases), the parameter $r_s/R_*$ must vary in order
to 
keep $\Delta M_{\rm env}$ fixed. From the ratio of the quantity $\Delta
M_{\rm 
env}$ in the perturbed state to the quantity $\Delta M_{\rm 
env}^{(0)}\equiv\Delta M_{\rm env}$ in the unperturbed state, together with 
equation (12), we obtain 

$$
{\lambda\over\lambda^{(0)}}={-\ln s-3(1-s)+{3\over 2}(1-s^2)-{1\over 3}(1-
s^3)\over-\ln s_0-3(1-s_0)+{3\over 2}(1-s_0^2)-{1\over
3}(1-s_0^3)},\eqno(16)
$$

\noindent where now $s\equiv r_s/R_*$, and $s_0$ is the value of $s$ in the 
unperturbed state.  If $r_s\sim 10^{10}$ cm and $R_*\sim 10^{12}$ cm -- \eg,
corresponding to a star leaving the AGB -- then $s_0\sim 10^{-2}$.  If we 
denote the temperature of the HBRS by $T_s$, we can use equation 
(13) to express it in the form of a ratio to the temperature $T_s^{(0)}$ in 
the unperturbed state: 
    
\def\sd{\strut\displaystyle} 

$$
y\equiv {T_s\over T_s^{(0)}}={\sd\left({1\over s}-1\right)\over\sd
\left({1\over s_0}-1\right)}.\eqno(17)
$$

\noindent Note that equations (16) and (17) provide a parametric relation
for 
the quantity $\lambda/\lambda^{(0)}$ in terms of the dimensionless
temperature 
$y$ and the constant $s_0$. 

    Next, consider the energy equation, using the form of $TdS$ that
includes 
both radiation and gas pressure.  The general expression can be written in
the 
form (\cf, Cox and Giuli 1968, p. 206 ff) 

$$
TdS=C_V\left[dT-{T\over \rho}(\Gamma_3 -1)d\rho\right].\eqno(18)
$$

\noindent Here the quantity $(\Gamma_3 -1)$ is the thermodynamic
derivative

$$
(\Gamma_3 -1)\equiv\left({\partial\ln T\over\partial\ln\rho}\right)_S =
{2\over
3}{(4-3\beta)\over(8-7\beta)},
$$

\noindent and the last equality gives the explicit expression for a mixture
of 
gas plus radiation.  Using equation (18), the energy equation (5e) becomes 

$$
{\partial L_r\over\partial r}=4\pi r^2\rho\left\{\epsilon
-C_V\left[{\partial
T\over \partial t}-{T\over \rho}(\Gamma_3 -1){\partial\rho\over\partial
t}\right] -vC_V\left[{\partial T\over \partial r}-{T\over \rho}(\Gamma_3 -
1){\partial\rho\over\partial r}\right]\right\}.\eqno(19)
$$

    We now wish to integrate this equation over the stellar envelope.  If we
assume 

$$
\epsilon=\epsilon_0 X\rho T^{\nu},\eqno(20)
$$

\noindent we can integrate the first term on the right-hand side of equation
(19) to obtain the luminosity generated by the HBRS,

$$
L_s\equiv\int_{r_s}^{R_*}\epsilon dM_r\equiv L_s^{(0)}
\left({X_s\over X_s^{(0)}}\right) \left({\rho_s\over \rho_s^{(0)}}\right)
\left({T_s\over T_s^{(0)}}\right)^{\nu}.\eqno(21a)
$$

\noindent Here the subscript $s$ denotes conditions at the HBRS
at 
some time $t$, and the superscript $(0)$ denotes the conditions in the 
unperturbed state.  The temperature exponent $\nu$ in the H-burning reaction
rate has the value $\nu\sim 2-6$ for the $pp$ chain, which is appropriate
for 
the relatively low shell-burning temperatures we consider here, while it has
the value $\nu\sim 10-20$ for the CNO cycle (\cf, Hansen and Kawaler 1994,
p. 238, 241).  From equations (11) and (12) we can also obtain the relative 
density distribution in the envelope, 

$$
{\rho_s\over\rho_s^{(0)}}={\lambda^{(0)}\over\lambda}\left({T_s\over
T_s^{(0)}}\right)^3,\eqno(22)
$$

\noindent and this can be used to replace $\rho_s/\rho_s^{(0)}$ in
equation
(21a).

    In the term involving time derivatives in equation (19), we use the 
continuity equation to eliminate $\partial\rho/\partial t$.  The resulting 
integrand involves the spatial derivative of the mass-flow rate $F$.  We 
assume that $F$ is independent of position as argued previously and 
accordingly neglect this term.  The remaining integral involving a time 
derivative can be written in the approximate form 

$$
\int_{r_s}^{R_*}C_V{\partial T\over \partial t}dM_r\approx C_V^*\Delta
M_{\rm
env}{dT_s\over dt},\eqno(21b)
$$

\noindent where $C_V^*$ is some average heat capacity for the envelope,
$\Delta 
M_{\rm env}$ is the envelope mass defined in equation (14), and $T_s$ is
again 
the temperature of the HBRS.  The remaining terms in equation
(19) 
represent heat advection.  We estimate them to be only $\sim 2\times
10^{-3}$ 
as large as other terms in this equation, and we have neglected them. 

    Collecting equations (21) together, we can thus write the integral of
equation (19) in the form

$$
L_*=L_1 + L_s^{(0)} \left({X_s\over X_s^{(0)}}\right)
\left({\lambda^{(0)}\over
\lambda}\right) \left({T_s\over T_s^{(0)}}\right)^{\nu+3} -
L_*^{(0)}\tau_{\rm
th}{d\over dt}\left({T_s\over T_s^{(0)}}\right),\eqno(23)
$$

\noindent where $L_1$ denotes the luminosity originating interior to the
HBRS, and

$$
\tau_{\rm th}\equiv{C_V^*\Delta M_{\rm env}T_s^0\over L_*^0}\sim 20\
{\rm years}\eqno(24)
$$

\noindent is the thermal relaxation timescale.  In evaluating $\tau_{\rm
th}$ 
we have taken $\Delta M_{\rm env}\sim 0.1\Msun$; $T\sim 10^6$ K, which may
be 
typical of a fuel-starved HBRS; and $L_*\sim 10^4\Lsun$. 

    In the final equation (5f), we first carry out a Galilean transformation
to a frame moving with the HBRS.  For any quantity $f(r,t)$, if
we 
define a new position coordinate to be $ r^{\prime}\equiv r-v_s t$, where 
$v_s$ is the propagation speed of the HBRS, the general relations
among the partial derivatives in the old and new frames are

$$
{\partial f\over\partial r^{\prime}}\biggr|_t  = {\partial f\over\partial
r}\biggr|_t \qquad {\rm and} \qquad{\partial f\over\partial t}
\biggr|_{r^{\prime}}  = {\partial f\over\partial t}\biggr|_r +
v_s{\partial
f\over\partial r}\biggr|_t.\eqno(25)
$$

\noindent If we apply the relations (25) we can express equation (5f) in
the frame moving with the HBRS:

$$
{\partial X\over\partial t}\biggr|_{r^\prime} + (v-v_s){\partial
X\over\partial {r^\prime}}\biggr|_t = -{\epsilon(r^{\prime},t)\over
Q^{\prime}},\eqno(26)
$$

\noindent where $Q^{\prime}$ is given by equation (1).  If we define the
position of the HBRS by the condition

$$
X(r_s,t)\equiv{1\over 2}X_0,\eqno(27)
$$

\noindent where $X_0\approx 0.7$ is the primordial H abundance, then at the 
location of this shell, we have $\partial X/\partial t|_{r_s}\equiv 0$. 
Evaluating equation (26) at this location thus gives the shell speed: 

$$
[v(r_s)-v_s]{\partial X\over\partial {r_s}}\biggr|_t = -
{\epsilon({r_s},t)\over Q^{\prime}},\eqno(28)
$$

\noindent where $v(r_s)$ is the speed of the flow, as measured in the rest 
frame of the star, at the position $r_s$ of the HBRS, and $v_s$
is 
the desired propagation speed of the shell.  If we use equation (28) to 
describe the presumed unperturbed steady-flow condition and subtract this
from 
equation (26), we obtain for the perturbed flow the result 

$$
{\partial X\over\partial t}\biggr|_{r^\prime} + (v-v_s){\partial
X\over\partial {r^\prime}}\biggr|_t - (v^{(0)}-v_s^{(0)}){\partial
X^{(0)}\over\partial {r^\prime}}\biggr|_t =
    -{[\epsilon(r^{\prime},t)-\epsilon^{(0)}(r^{\prime},t)]\over
Q^{\prime}},\eqno(29)
$$

    We now need to work out the integrals of the various terms in this
equation.  The first term can be written in the form

$$
\int_{M_s}^{M_*}{\partial X\over\partial t}dM_{r^{\prime}} \equiv \Delta
M_{\rm nuc}\overline{\partial X\over\partial t}\approx \Delta M_{\rm
nuc}{dX_s\over dt}.\eqno(30a)
$$

\noindent This equation defines the spatial average of the time derivative, 
which we approximate as the time derivative of the H abundance in the
HBRS.  Note that $X$ is constant in time and space except in the very thin
HBRS.  Accordingly the mass $\Delta M_{\rm nuc}$ in 
HBRS regions is relatively small; we use the approximate value 
$\Delta M_{\rm nuc}\sim 5\times 10^{-4}\Msun$ for this mass. 
(This value is actually consistent with 
the final H-burning shell mass found by Iben (1984), Iben and MacDonald 
(1986), and Herwig \etal\ (1999) in late AGB or immediate post-AGB phases of
stellar evolution. However since our model also 
requires a significant luminosity internal to the HBRS, $L_1$, 
we do not require that the HBRS is the entire H-burning shell.  Some arbitrary 
ratio of H and He could be burning below the HBRS to provide $L_1$.)  

In the terms in equation (29) that involve spatial 
gradients of $X$, we write $dM_{r^{\prime}}=4\pi 
(r^{\prime})^2\rho(r^{\prime},t)dr^{\prime}$ and use equation (6) to replace
$v$ with the mass-flow rate $F$, which we assume to be independent of 
position.  The resulting integral has the form 

$$
\int_{M_s}^{M_*}\left[(v-v_s){\partial X\over\partial {r^\prime}} -
(v^{(0)}-
v_s^{(0)}){\partial X^{(0)}\over\partial {r^\prime}}\right]dM_{r^{\prime}}
=
[(F-F^{(0)})-(F_s-F_s^{(0)})]X_0,\eqno(30b)
$$

\noindent where $F_s\equiv 4\pi r^2\rho v_s$, and $X_0$ is again the
primordial 
H abundance.  The integral of the terms on the right-hand side of equation 
(29) just gives the difference in the HBRS luminosities in the 
perturbed and unperturbed states.  Collecting these results together, and
for 
simplicity neglecting $F_s-F_s^{(0)}$, we write the integral of the $X$ 
equation in the approximate form 

$$
\Delta M_{\rm nuc}{dX_s\over dt} +(F-F^{(0)})X_0=-{(L_s-L_s^{(0)})\over
Q^{\prime}}.\eqno(31)
$$

    The system of equations describing the presumed relaxation oscillations 
now consists of (23) and (31), with the dimensionless luminosity $\lambda$ 
expressed in terms of the dimensionless shell temperature $y$ by the 
parametric equations (16) and (17).  For given values of $y$ and $s_0$, we 
solve equation (17) to obtain $s$, and we use these values in equation (16)
to 
obtain $\lambda/\lambda^{(0)}$.  For convenience, we collect the resulting 
equations together here, defining $x\equiv X_s/X_s^{(0)}$: 

$$
L_*=L_1 + L_s^{(0)} \left({\lambda^{(0)}\over \lambda}\right) xy^{\nu+3} -
L_*^{(0)}\tau_{\rm th}{dy\over dt},\eqno(32a)
$$

$$
X_s^{(0)}\Delta M_{\rm nuc}{dx\over dt} +(F-F^{(0)})X_0=-{L_s^{(0)}\over
Q^{\prime}}\left[\left({\lambda^{(0)}\over \lambda}\right) xy^{\nu+3}-
1\right],\eqno(32b)
$$

$$
\left({1\over s}-1\right)=y\left({1\over s_0}-1\right),\eqno(32c)
$$

\noindent and

$$
{\lambda\over\lambda^{(0)}}={-\ln s-3(1-s)+{3\over 2}(1-s^2)-{1\over 3}(1-
s^3)\over-\ln s_0-3(1-s_0)+{3\over 2}(1-s_0^2)-{1\over
3}(1-s_0^3)}.\eqno(32d)
$$

    For radiation-driven mass loss, we assume the simple form given by
equation (4), which we write as

$$
\dot M=-\alpha L_*.\eqno(33)
$$

\noindent Combining (6), (9), and (33), we obtain

$$
{\dot M\over\dot M^{(0)}}={F\over F^{(0)}}={L_*\over
L_*^{(0)}}={\lambda\over \lambda^{(0)}}.\eqno(34)
$$

\noindent Equation (34) eliminates equation (6) and expresses the mass flux 
$F$ in terms of $\lambda$. We can now put the equations in fully
dimensionless 
form.  Dividing equation (32a) by $L_*^{(0)}$, we obtain 

$$
{\lambda\over\lambda^{(0)}}=A+ (1-A)\left({\lambda^{(0)}\over
\lambda}\right)
xy^{\nu+3} - \tau_{\rm th}{dy\over dt},\eqno(35a)
$$

\noindent where $A\equiv L_1/L_*^{(0)}$.  Note that, in a truly steady state
with $dy/dt=0$, we have $x=1$, $y=1$, and $\lambda=\lambda^{(0)}$, obtaining
from equation (23) $L_*^{(0)}\equiv L_s^{(0)}+L_1$.  Similarly, we divide 
equation (32b) by the quantity $X_s^{(0)}\Delta M_{\rm nuc}$ and use
equation 
(34) to obtain 

$$
{dx\over dt} +\left({\lambda\over\lambda^{(0)}}-1\right){1\over\tau_{\rm
fsm}}=-{1\over\tau_{\rm ex}}\left[\left({\lambda^{(0)}\over
\lambda}\right)
xy^{\nu+3}-1\right].\eqno(35b)
$$

\noindent In equations (35), $\tau_{\rm th}$ is given by equation (24).  We 
define

$$
\tau_{\rm fsm}\equiv{X_s^{(0)}\Delta M_{\rm nuc}\over X_0F^{(0)}} 
\equiv{X_s^{(0)}\Delta M_{\rm nuc}\over X_0|\dot M|}
    \sim 25\ {\rm years},\eqno(36a) 
$$

\noindent as the timescale on which the fuel supply is modulated by mass 
loss; \ie, the time required for mass loss to strip material out of the 
HBRS.  In evaluating $\tau_{\rm fsm}$ we assume the mass in the 
HBRS to be $\Delta M_{\rm nuc}\sim 5\times 10^{-4}\Msun$, as
noted 
above, and we assume $\dot M\sim -10^{-5}\Msun$ yr$^{-1}$ to be the
mass-loss 
rate.  We define the fuel-exhaustion timescale similarly: 

$$
\tau_{\rm ex}\equiv {Q^{\prime}X_s^0\Delta M_{\rm nuc}\over L_s^0}\sim
1.75\times 10^4\ {\rm years}.\eqno(36b)
$$

    An important constraint is provided by the fact that the H abundance can
never exceed the primordial value $X_0$ nor become negative.  From equation 
(27), this correspondingly constrains our dimensionless parameter $x\equiv 
X_s/X_s^{(0)}$ to the range 

$$
0\leq x\leq (X_0/{1\over 2}X_0)=2.\eqno(37)
$$

\noindent Equations (35a) and (35b), together with the constraint equation 
(37), the parametric relations (32c) and (32d), and the definitions of the 
characteristic timescales (24), (36a), and (36b), constitute the system of 
equations that characterize this relaxation-oscillator model.  In these 
equations, the quantity $\lambda^{(0)}$ is a dimensionless number presumably
just slightly less than unity (\ie, the luminosity is just slightly less
than the Eddington luminosity).

The mathematical nature of the governing system of nonlinear
differential equations (35a,b) is better seen by writing them in the
fully nondimensional form 

$$
{dy\over dt} = A+(1-A)xh(y) - g(y), \eqno(38a)
$$

$$
{dx\over dt} = -\gamma_1[g(y)-1] - \gamma_2[x h(y) - 1], \eqno(38b)
$$

\noindent where 

$$
g(y)\equiv{\lambda\over\lambda^{(0)}},\qquad  h(y)\equiv{y^{(\nu
+3)}\over g(y)},
\qquad \gamma_1 \equiv {\tau_{\rm th}\over\tau_{\rm fsm}}, \qquad \gamma_2 \equiv 
{\tau_{\rm th}\over\tau_{ex}},\eqno(39)
$$

\noindent and where $t$ is now a nondimensional time, scaled by the
thermal time scale $\tau_{\rm th}$. Note that  $g(y)\equiv
\lambda/\lambda^{(0)}$ is a nonlinear function of $y$ defined by
equations (32c) and (32d). Note also that $g(1)=h(1)=1$ and that equations
(38)  have the equilibrium solution $x=y=1$. A linear
stability analysis, presented in the Appendix, shows that this 
equilibrium is unstable to growing oscillations for a range of 
values of the stellar parameters.

\vskip 10pt

\noindent{\bf 5. Numerical Solutions of the Model Equations}

\vskip 10pt

We have carried out a number of numerical solutions of the nonlinear 
system (38) in order to explore the behavior of the relaxation 
oscillations and to determine their dependence on the various 
parameters in the model.
As a specific example, consider first a benchmark model with $M_*=0.7\Msun$, 
$L_*=10^4\Lsun$, $R_*=10^{13}$ cm, $\Delta M_{\rm env}=0.1\Msun$, $\Delta 
M_{\rm nuc}=5\times 10^{-4}\Msun$, and $T_s^{(0)}=10^6$ K.  For these 
parameters, equation (13) and the definition $s_0\equiv r_s^{(0)}/R_*$ yield
$s_0=0.03$.  We adopt the values  $\nu=3$ and $A\equiv
L_1/L_*^{(0)}=0.9$ in this model.
These values yield $\tau_{\rm th}=20$ years, $\tau_{\rm fsm}=25$ years, and  
$\tau_{\rm ex}=1.75\times 10^4$ years.  For these parameter choices, the 
numerical solution of equations (38) yields sustained relaxation
oscillations with a period $P\approx 72\tau_{\rm th}\approx$ 1400 years, as shown in
Fig. 1.  During the initial rise in temperature and luminosity, the H abundance
$X$ drops rapidly, reaching $X=0$ in $\sim 39\tau_{\rm th}\approx$ 780 years.
The resulting relaxation oscillations reach a peak temperature $T_s\approx
6\times 10^7$ K and a peak luminosity $L_*\approx 3.28L_*^{(0)}$.  The half-width of
a pulse, defined as the time during which the luminosity exceeds half its peak
value, is $\Delta t\sim 28\tau_{\rm th}\approx$ 560 years $=0.39 P$.  We 
define the \lq\lq duty cycle\rq\rq\ $W$ of a pulse as the time required for 
the shell temperature to relax from its peak value to $T_s^{(0)}$ (\ie, for 
$y$ to return to the value $y=1$).  This is approximately the time required 
for the star to radiate away the heat generated in the thermonuclear shell 
flash.  A simple analytic argument based on our dimensionless model
equations gives the duty cycle as $W\sim 26\tau_{\rm th}$, while our explicit
numerical calculation yields $W=39.4\tau_{\rm th}=0.55 P$. 

    The H abundance in the shell remains vanishingly small until the
luminosity 
-- and the accompanying mass-loss rate -- has declined sufficiently for the 
high-temperature shell again to overtake the H fuel supply.  The recovery 
between successive pulses is governed by the time required for the H
abundance 
-- and the associated energy-generation rate -- to recover to, and then 
exceed, the equilibrium rate.  The time between pulses is $P\approx 
72\tau_{\rm th}=1400$ years, closely comparable to the timescale between
the observed circumnebular shells in NGC 6543.  We obtain similar sustained 
relaxation oscillations starting from shell temperatures $T_s$ either
slightly less than or slightly greater than $T_s^{(0)}$, where $T_s^{(0)}$ is the 
presumed steady-state burning temperature; in this sense, these results are 
robust. 

    We have investigated the numerical accuracy of these results in two
ways.  Calculations with timesteps selected to be a factor of two larger and
smaller than the values used in the calculations we report here give
values of $P$ within $\pm 0.2$\%, values of $W$ within $\pm 1.8$\%,
and values of the peak temperature within $\pm 2.0$\%.  Calculations
for this same model obtained using two different algorithms yield
values of $P$ that agree to within 20\%, values of $W$ within 20\%,
and values of the peak temperature within about 3\%. 

    We have explored the variations in the properties of these relaxation 
oscillations for the range of model parameters listed in Table 1 below.
Model F is the illustrative benchmark case discussed above.  The
numerical results are summarized in Table 2 below.  Depending upon the
specific choice of parameters, we obtain oscillations with a range of
periods, duty cycles, and peak temperatures and luminosities.  Note
that the peak temperatures for all except model F exceed $10^8$ K,
violating our simplifying model assumption that energy production in
the shell is due solely to the H-burning $pp$ reactions. 

\vskip 10pt

$$
\table
\tablespec{\r\r\r\l\r\r\r\r\r}
\body{
\header{Table~1. Model Parameters$^*$}
\skip{5pt}
\hline
\skip{2pt}
\hline
|Model|$-\dot M$|$T_s^{(0)}$|$s_0$|$\nu$|$A$|
    $\tau_{\rm th}$|$\tau_{\rm fsm}$|$\tau_{\rm ex}$|\end
\hline
|A|$10^{-6}$|$10^7$|0.003|3|0.9|200|500|$3.50\times 10^4$|\end
\hline
|B|$10^{-5}$|$10^7$|0.003|3|0.9|200| 50|$3.50\times 10^4$|\end
\hline
|C|$10^{-6}$|$10^6$|0.03 |3|0.9| 20|500|$3.50\times 10^4$|\end
\hline
|D|$10^{-5}$|$10^6$|0.03 |3|0.9| 20| 50|$3.50\times 10^4$|\end
\hline
|E|$10^{-4}$|$10^6$|0.03 |3|0.9|  2|  5|$3.50\times 10^4$|\end
\hline
|F|$10^{-5}$|$10^6$|0.03 |3|0.9| 20| 25|$1.75\times 10^4$|\end
\hline
}
\endtable
$$

\item{$^*$}The mass-loss rate $\dot M$ is in $\Msun$ yr$^{-1}$, $T_s^{(0)}$ 
is in K, and all timescales are in years. 

\vskip 10pt

$$
\table
\tablewidth{5.0truein}
\tablespec{\r\r\r\r\r}
\body{
\header{Table~2. Model Results$^*$}
\skip{5pt}
\hline
\skip{2pt}
\hline
|Model|$P$|$T_s^{\rm max}/T_s^{(0)}$|$L_s^{\rm max}/L_s^{(0)}$|$W$|\end
\hline
|A|$2.28\times 10^4$|  31.7|1.86|0.45|\end
\hline
|B|$1.38\times 10^4$|  30.3|1.85|0.67|\end
\hline
|C|$9.12\times 10^3$| 171.7|3.85|0.16|\end
\hline
|D|$2.47\times 10^3$| 140.9|3.73|0.55|\end
\hline
|E|$9.29\times 10^2$|1,409.|5.05|0.92|\end
\hline
|F|$1.44\times 10^3$|  62.2|3.28|0.55|\end
\hline
}
\endtable
$$

\item{$^*$}The oscillation period $P$ is in years, and the \lq\lq duty 
cycle\rq\rq\ $W$ of the pulse is defined as the fraction of the pulse period
required for the shell temperature $T_s$ to decline from its maximum value 
$T_s^{\rm max}$ to the mean steady-state temperature $T_s^{(0)}$.

\vskip 10pt

    In order to investigate the dependence of the oscillations on 
the paramter $A\equiv L_1/L_*^{(0)}$ we did a series of calculations 
using $\nu=3$ and $\Delta M_{\rm nuc}=0.001$. In this case we find 
that oscillations occur only when the luminosity ratio 
$A\equiv L_1/L_*^{(0)}$ lies in the range $0.84<A<0.92$. We can 
also estimate the range of values of $A$ that give
oscillations from a linear stability analysis of the governing 
nonlinear equations (see the Appendix). In this case, the linear 
analysis predicts growing oscillations in the range 
$0.854<A<0.907$.  We have not found relaxation 
oscillations for $A\leq 0.8$, even though we have carried out a number
of calculations in this range.  For example, with $A=0.3$ and 
$\nu =10$, the system goes through a single flash and then stabilizes at an 
HBRS luminosity lower than the initial assumed steady-state 
luminosity.  Evidently this is not a self-consistent solution of the
problem. Further, neglect of the luminosity $L_1$ emerging from below
the HBRS 
(\ie, setting $A=0$) prevents matter in the envelope from re-heating 
when the luminosity drops and mass loss shuts down.  Conversely, if the 
(presumed constant) interior luminosity $L_1$ is too large, the system goes 
through a single thermonuclear pulse and then settles down into a state of 
steady burning.  For example, with $L_1/L_*^{(0)}=0.95$, $\nu=3$, $\tau_{\rm
fsm}/\tau_{\rm th}=2.5$, and $\tau_{\rm ex}/\tau_{\rm th}=3000$, there is no
oscillation, and the shell temperature and luminosity stabilize at the
assumed steady-state values.  Thus, there appears to be a rather small region of 
parameter space in which the star exhibits such fuel-supply-limited
relaxation oscillations. 

    We have studied the dependence of the relaxation oscillations upon the 
assumed value of $\Delta M_{\rm nuc}$, taking $A=0.9$, $\nu=3$, and
$s_0=0.01$ for all models.  As shown in Table 3 below, we found that
the oscillation 
period $P$ decreases and the peak values of the temperature and luminosity 
decline as $\Delta M_{\rm nuc}$ decreases.  The period appears to approach
an asymptotic value $\sim (15-17)\times \tau_{\rm th}$ for small HBRS 
masses.  For the model in Table 3 with the lowest listed value of $\Delta 
M_{\rm nuc}$, the amount of energy available from the shell flash is so
small that the oscillations actually decay slowly in time, rather than
attaining a roughly constant amplitude. 

\vskip 10pt

$$
\table
\tablewidth{5.0truein}
\tablespec{\l\l\l\l}
\body{
\header{Table~3. Variations with $\Delta M_{\rm nuc}$}
\skip{5pt}
\hline
\skip{2pt}
\hline
|$\Delta M_{\rm nuc}/\Msun$|$P/\tau_{\rm th}$|
    $T_s^{\rm max}/T_s^{(0)}$|$L_s^{\rm max}/L_s^{(0)}$|\end
\hline
|0.001    |33|2.8|1.36|\end
\hline
|0.0006   |26|2.1|1.26|\end
\hline
|0.0005   |22|1.8|1.20|\end
\hline
|0.0004   |20|1.6|1.16|\end
\hline
|0.0003333|19|1.5|1.14|\end
\hline
|0.0003077|18|1.25|1.08|\end
\hline
|0.0002857|17--15|1.1|1.03|\end
\hline
}
\endtable
$$

\vskip 10pt

    We have also studied the dependence of the relaxation oscillations upon 
the assumed value of $\nu$, taking $\Delta M_{\rm nuc}=0.001\Msun$, $A=0.9$, 
$T_s^{(0)}=10^7$ K, and $s_0=0.01$ for all models.  As shown in Table 4
below, 
the system does not depend very much upon $\nu$, apparently because the 
temperature spikes so rapidly.  We found no oscillations for $\nu\gsim 7$.  
Within the limitations of our simplified model, a stronger dependence of the
thermonuclear rate upon $T$ apparently liberates energy so rapidly that the 
burning simply adjusts to a different stable-burning state.  For 
calculations with $\nu=6$, the peak temperatures reach $\sim 5\times 10^8$
K, 
and our approximation for the nuclear burning rate is invalid. 

\vskip 10pt

$$
\table
\tablewidth{5.0truein}
\tablespec{\l\l\l\l}
\body{
\header{Table~4. Variations with $\nu$}
\skip{5pt}
\hline
\skip{2pt}
\hline
|$\nu$|$P/\tau_{\rm th}$|
    $T_s^{\rm max}/T_s^{(0)}$|$L_s^{\rm max}/L_s^{(0)}$|\end
\hline
|3|33|2.8|1.36|\end
\hline
|4|35|3.5|1.44|\end
\hline
|5|38|3.9|1.47|\end
\hline
|6|51|3.9|1.48|\end
\hline
}
\endtable
$$

\vskip 10pt

One aspect of the results surprised us:
the conditions under which we found oscillations 
correspond to very low values of the nuclear-energy-generation exponent
$\nu$. 
The values for which we find oscillations correspond to energy generation by
the $pp$ chain, whereas we had expected that the CNO cycle would dominate. 
Conceivably, this might also be a consequence of the terminal stage of shell
burning. 

    Finally, we note that in order to justify our suggested link between 
these relaxation oscillations and ejection of mass shells, the time
scale for acceleration
from rest to the escape velocity must be less than the oscillation
period. This condition is easily satisfied; the acceleration time for material
illuminated from below by luminosity $L_* \gsim L_{\rm Edd}$ is $\sim 0.13\ {\rm 
year}\times  (M_*/M_{\sun})^{-1/2} (R_*/10^{13}{\rm cm})^{3/2}(L_*/L_{\rm
Edd} -  1)^{-1}$, assuming the Thomson cross section as a conservative bound.
This acceleration time
is always much less than the $\gsim 10^3$ year oscillation period we find,
so that we are well justified in applying our model to PN systems with evidence
for periodic ejecta. 

\vskip 10pt

\noindent{\bf 6. Conclusions}

\vskip 10pt

    We have shown that fuel-supply-limited relaxation oscillations may
develop 
in some stars near the ends of their nuclear-active lifetimes just prior to 
the ejection of a planetary nebula.  In the benchmark model we discuss in this paper, 
with $M_*\lsim \Msun$, $L_*\sim 10^4\Lsun$, and $|\dot M|\sim 10^{-5}\Msun$ 
yr$^{-1}$, such oscillations occur when the luminosity generated by the
H-burning relaxation shell (HBRS) 
falls to $\sim$ 10\% of the total luminosity of the star.
The pulsation timescale we find to be $\sim 72\tau_{\rm th}\approx 57 \tau_{\rm 
fsm}\sim 1400$ years, closely comparable to the timescale inferred for the 
observed circumnebular shells (Balick et al. 2001; Zijlstra et al. 2002), 
for the present choice of parameters.  While 
our choice of parameters does produce results consistent with the observed 
circumnebular shells, it is not clear whether such parameters are actually 
attained during the evolution of a real star.  To determine whether such 
oscillations occur in more realistic stellar models, or whether they
do indeed provide the correct explanation for the observed shells
surrounding some planetary nebulae, will require much more detailed 
calculations of stellar evolution with mass loss than have been done to date. 

\vskip 10pt

\noindent{\bf Acknowledgments}

\vskip 10pt

    HMVH acknowledges support from the National Science Foundation, which
made his participation in this work possible.  He also wishes to express his 
gratitude to the University of Rochester and to the Department of
Terrestrial Magnetism at the Carnegie Institution of Washington, both of which
graciously served as host institutions during the course of this work.
In particular, he thanks S. Keiser for her help with the calculations
and display of the results.  JHT acknowledges support from NSF grant 
AST95-28398 and thanks Clare Hall and the Department of Applied
Mathematics and Theoretical Physics, University of Cambridge, for
their hospitality during the final preparation of this manuscript.  
EB acknowledges support from DOE  
grant DE-FG02-00ER54600, and the Laboratory for Laser Energetics. 
Support for AF was provided at the University of Rochester by NSF 
grant AST-9702484, AST-0098442, NASA grant NAG5-8428 and the 
Laboratory for Laser Energetics. 

\vskip 10pt

\noindent{\bf Figure Caption}

\vskip 10pt

\noindent{\bf Fig. 1.} Fuel-supply-limited relaxation oscillations of the
model described in \S 4 (model F in Tables 1 and 2).  
The dotted curve represents the dimensionless
temperature $y\equiv T_s/T_s^{(0)}$, the dashed curve gives the dimensionless 
HBRS luminosity $L_s/L_s^{(0)}$, and the solid curve is the 
dimensionless H abundance $x\equiv X_s/X_s^{(0)}$ in the HBRS. The 
abscissa gives the time elapsed from the start of the calculation in units
of the thermal timescale defined by equation (24), $t/\tau_{\rm th}$.  At the 
onset of a pulse, the rapid increase in luminosity drives the H abundance in
the HBRS to zero in less than $\sim 39\tau_{\rm th}\approx 780$ 
years. During this initial phase, the shell temperature rises rapidly to a 
peak value $\sim 62$ times the intial value.  The energy generated during
this pulse is gradually radiated away during the next $\sim 39\tau_{\rm th}\sim 
780$ years, and the luminosity declines toward the initial luminosity on the
same timescale.  When the surface luminosity and the associated mass-loss
rate have declined sufficiently, the H abundance in the HBRS begins to 
build up again.  In $\sim 72\tau_{\rm th}\sim 1400$ years $\sim 57\tau_{\rm
fsm}$, the energy generation rate in the HBRS builds up to a
level exceeding the level of steady-state burning, and another pulse occurs. 

\vskip 10pt

\noindent{\bf Appendix: Linear Stability Analysis}

\vskip 10pt

To investigate the stability of the equilibrium solution $x=y=1$ of
equations (38), let $x=1+\xi$ and $y=1+\eta$ and expand the
functions $g(y)$ and $h(y)$ in Taylor series around $y=1$ in the form 

$$
g(1+\eta)=1+b \eta + {\rm O}(\eta^2), \qquad h(1+\eta)=1+c \eta
+ {\rm O}(\eta^2),\eqno(A1)
$$

\noindent where the coefficient $b$ depends on the parameter $s_0$
and the coefficient $c$ depends on the parameters $s_0$ and 
$\nu$. Substituting these expressions into equations (38) and 
linearizing yields the equations

$$ 
{d\xi\over dt}=- \gamma_2 \xi - (\gamma_1 b + \gamma_2
c)\eta,\eqno(A2a)
$$ 

$$
{d\eta\over dt}=(1-A)\xi + [c(1-A)-b]\eta.\eqno(A2b)
$$

\noindent If we assume a nonzero solution of these equations in the form 
$\xi=\xi_0 {\rm exp}(\alpha t)$, $\eta=\eta_0 {\rm exp}(\alpha t)$, 
then we find that $\alpha$ satisfies a quadratic equation with roots

$$
\alpha = [B \pm (B^2-4C)^{1/2}]/2,\eqno(A3)
$$

\noindent where

$$
B=(1-A)c-b-\gamma_2, \qquad C=[(1-A)\gamma_1 + \gamma_2]b.\eqno(A4)
 $$ 

\noindent The solution is an oscillation when $B^2-4C < 0$ and these 
oscillations are growing when $B > 0$.

Using this analysis, we can determine the regions of parameter space
in which we expect to find growing oscillations. For example, with 
$s_0=0.03$, $\nu=3$, $\gamma_1=0.8$, and $\gamma_2=0.001143$ (as in 
model F in Table 1), we find that the solutions of the linearized 
equations are oscillatory for values of the parameter $A$ in the 
range $0.807<A<0.957$, and these oscillations are growing  
for $A$ in the range $0.807 < A < 0.908$.

\vskip 10pt

\noindent{\bf References}

\vskip 10pt

\item{}Balick, B., Wilson, J., and Hajian, A. R. 2001, {\it A. J.}, {\bf
121}, 354. 

\item{}Blackman, E.G., Frank, A., Markiel, J.A., Thomas, J.H., and Van
Horn, H.~M.\ 2001, Nature 409, 485

\item{}Bowen, G. H. 1988, {\it Ap. J.}, {\bf 329}, 299.

\item{}Bowen, G. H., and Willson, L. A. 1991, {\it Ap. J.}, {\bf
375}, L53. 

\item{}Clayton, D. D. 1968, {\it Principles of Stellar Evolution and
Nucleosynthesis} (McGraw-Hill: New York).

\item{}Cox, A. 2000, ed. {\it Allen's Astrophysical Quantities},
4$^{\rm th}$ ed. (Springer-Verlag: New York).

\item{}Cox, J. P., and Giuli, R. T. 1968, {\it Principles of Stellar
Structure} (Gordon and Breach: New York).

\item{}Dorman, B., Rood, R. T., and O'Connell, R. W. 1993, {\it Ap. J.},
{\bf 419}, 596. 

\item{}Garcia-Segura \etal\ 2002 {\tt *** Need refc. ***}

\item{}Hansen, C. J., and Kawaler, S. D. 1994, {\it Stellar Interiors:
Physical Principles, Structure, and Evolution}, (Springer-Verlag: New
York).

\item{}Herwig, F., Bl\"ocker, T., Langer, N., and Driebe, T. 1999, {\it A.
\& 
A.}, {\bf 349}, L5.

\item{}Iben, I., Jr. 1984, {\it Ap. J.}, {\bf 277}, 333.

\item{}Iben, Jr., I. 1987, in {\it Late Stage of Stellar Evolution}, ed. S. 
Kwok and S. R. Pottasch (Reidel: Dordrecht), p. 175. 

\item{}Iben, Jr., I. 1991, {\it Ap. J. Suppl.}, {bf 76}, 55.

\item{}Iben, I., Jr., and MacDonald, J. 1986, {\it Ap. J.}, {\bf 301}, 164.

\item{}Iben, Jr., I., and Renzini, A. 1983, {\it Ann. Rev. Astr. Ap.},
{\bf 21}, 271. 

\item{}Koesterke, L., Dreizler, S., and Rauch, T. 1998, {\it A. \& A.},
{\bf 330}, 1041. 

\item{}Koesterke, L., and Werner, K. 1998, {\it Ap. J.}, {\bf 500}, L59. 

\item{}Landau, L. D., and Lifshitz, E. M. 1959, {\it Fluid Mechanics},
(Pergamon Press: Oxford).

\item{}Mastrodemos, N., and Morris, M. 1998, {\it Ap. J.}, {\bf 497}, 303.

\item{}Mastrodemos, N., and Morris, M. 1999, {\it Ap. J.}, {\bf 523}, 357.

\item{}Mauron, N., and Huggins, P. J. 2000, {\it Ap. J.}, {\bf 359}, 707.

\item{}Mazzitelli, I., and D'Antona, F. 1986, {\it Ap. J.}, {\bf
308}, 706.

\item{}Zijlstra, 
A.A., Bedding, T.R., \& Mattei, J.A.\ 2002, MNRAS, accepted 
for publication, astro-ph/0203328

\item{}Perinotto, M. 1989, in {\it Planetary Nebulae}, ed. S.
Torres-Peimbert (Kluwer?: Dordrecht), p. 293. 

\item{}Reimers, D. 1975a, {\it Mem. Soc. R. Sci. Liege, 6 Ser.}, {\bf 8},
369.

\item{}Reimers, D. 1975b, in {\it Problems in Stellar Atmospheres and
Envelopes}, ed. B. Baschek, W. H. Kegel, and G. Traving (Springer:
Berlin?), p. 229. 

\item{}Renzini, A. 1981, in {\it Physical Processes in Red Giants}, ed. I.
Iben, Jr., and A. Renzini (Reidel: Dordrecht), p. 431.

\item{}Sch\"onberner, D. 1983, {\it Ap. J.}, {\bf 272}, 708.

\item{}Simis, Y., Icke, V., and Dominik, C. 2001,
http://xxx.lanl.gov/abs/astro-ph/0103347.

\item{}Soker and Harpaz 1988 {\tt *** Need refc. ***}

\item{}Terzian, Y., and Hajian, Y. 2000, in {\it Asymmetric Planetary
Nebulae. 
II. From Origins to Microstructure}, ed. J. H. Kastner, N. Soker, and S. A. 
Rappaport, {\bf CS-199}, (Astron. Soc. Pacific: San Francisco), p. 33.

\item{}Vassiliadis, E., and Wood, P. R. 1993, {\it Ap. J.}, {\bf 413},
641.

\item{}Willson, L. A. 2000, {\it Ann. Revs. Astron. Ap.}, {\bf 38}, 573.

\vfill
\eject
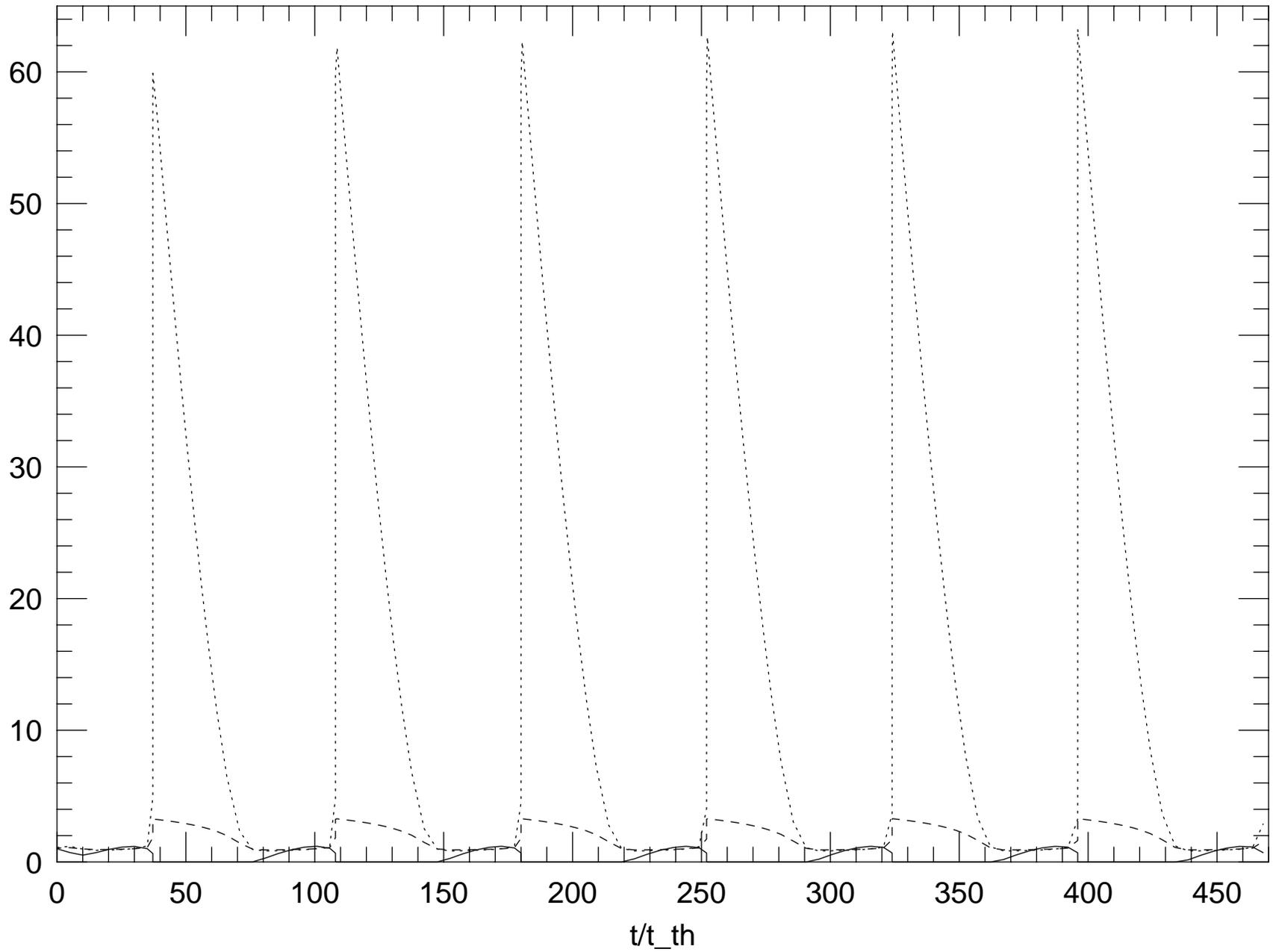
\bye